\documentclass[aps,prb,amsmath,amssymb,reprint]{revtex4-2}
\usepackage{graphicx} 
\usepackage{newtxtext} 
\usepackage{newtxmath} 
\usepackage{bm} 
\usepackage[T1]{fontenc} 
\usepackage{physics}
\usepackage{amsmath} 
\usepackage{siunitx} 

\usepackage[colorlinks,linkcolor=blue,anchorcolor=blue,citecolor=blue]{hyperref}
\hypersetup{
    colorlinks=true,
    linkcolor=blue,
    anchorcolor=blue,
    citecolor=blue,
    urlcolor=blue
}

\begin{document}

\preprint{APS/123-QED}

\title{Effect of Spin-Orbit Coupling on Anomalous Quantum Oscillations in InAs/GaSb Quantum Wells}

\author{Xinlong Du$^{1}$, Chao Wang$^{2}$, Bo Ying$^{1}$, and Juntao Song$^{1}$}
\email{jtsong@hebtu.edu.cn}
\affiliation{$^1$College of Physics and Hebei Advanced Thin Films Laboratory, Hebei Normal University, Shijiazhuang, Hebei 050024, China \\
$^2$College of Physics, Shijiazhuang University, Shijiazhuang, Hebei 050035, China \\
}

\date{\today}% It is always \today, today,
             %  but any date may be explicitly specified2
\begin{abstract}
We theoretically study the effect of spin-orbit coupling (SOC) on anomalous quantum oscillations in InAs/GaSb quantum wells. By comparing different cases, we show that SOC induces two opposing effects on anomalous quantum oscillations: it suppresses the oscillations in the clean case, while enhancing them in the disordered case. Using an effective model, we analyze in detail the origins of anomalous oscillations in both clean and disordered cases. Based on these origins, we explain why SOC suppresses or enhances the anomalous oscillations in different cases, thereby extending the understanding of the conventional theory. Moreover, in the disordered case, SOC can induce a phase shift of the anomalous oscillations. We further identify a parameter window where the anomalous oscillations are significantly enhanced in the presence of both disorder and SOC. These results provide a theoretical basis for understanding the role of SOC in anomalous quantum oscillations. 
\end{abstract}

\maketitle
\section{\label{sec:level1}Introduction}
Quantum oscillation is a unique manifestation of Landau quantization and provides crucial insights into the electronic properties near the Fermi surface \cite{01,002,003}. In metals, the application of a strong magnetic field would quantize the energy levels into Landau levels (LLs). As the magnetic field varies, these LLs cross the Fermi level periodically, leading to oscillations of the density of states (DOS). The DOS oscillations give rise to the oscillatory behavior of various physical quantities, such as the magnetic susceptibility (de Haas-van Alphen effect, dHvA) and the resistivity (Shubnikov-de Haas effect, SdH) \cite{02,03}. Quantum oscillations have long been regarded as an exclusive phenomenon of metals \cite{04,05}.

However, this canonical understanding of quantum oscillations has been challenged by recent observations of the Shubnikov-de Haas and de Haas-van Alphen effects in certain insulating systems, including the Kondo insulators SmB$_6$ \cite{06} and YbB$_{12}$ \cite{07,08}; the topological excitonic insulator WTe$_2$ \cite{09}; and the quantum spin liquid candidate $\alpha$-RuCl$_3$ \cite{10}. These anomalous quantum oscillations have motivated extensive theoretical studies, such as Kondo breakdown \cite{11}, the narrow hybridization gap model \cite{12,13}, disorder effects \cite{14,15}, and electron--hole interactions \cite{16,17,18,19,20,21,22,23}. Among these, the heavy-fermion model has been widely used to explain the anomalous oscillations observed in SmB$_6$. Although it provides a valuable theoretical framework, the model still has some limitations. Therefore, it is necessary to find a theoretical model that accurately represents realistic materials and captures phenomena observable in experiments.

The InAs/GaSb quantum well, a topological insulator, is an ideal platform for studying anomalous quantum oscillations \cite{26,27,28,29,30,31,32,33}. Owing to its type-II heterostructure, the InAs/GaSb quantum well exhibits an inverted band structure \cite{26}. Its dispersion and bandgap are highly tunable through the well thickness or an applied electric field \cite{27,28}. In addition, the system hosts a sizable Rashba spin–orbit coupling (SOC) arising from broken structural inversion symmetry, which further modifies the band structure \cite{29,30}. Compared to two-band heavy-fermion models, the InAs/GaSb quantum well incorporates material parameters closer to experimental conditions, enables tunable SOC strength, and provides flexibility for studying impurity doping. Recent transport experiments have revealed persistent quantum oscillations in the insulating phase of this system \cite{34,35,36}, further indicating that it is a suitable platform for studying the mechanisms of anomalous quantum oscillations.

In the InAs/GaSb system, the effects of disorder are particularly important, as they influence the physical properties of the material. Specifically, the quantum transport properties of this system are highly sensitive to disorder \cite{37,38,39,40,41}. For instance, in the HgTe/CdTe system, which is similar to the InAs/GaSb system, disorder can modify the bulk bandgap, alter conductivity, and even drive a transition from a trivial insulator to a topological Anderson insulator \cite{42,43,44,45,46,47,48}. Moreover, disorder-induced in-gap states have been used in both theoretical and experimental studies to account for anomalous quantum oscillations \cite{14,15,35}.

In addition to the effects of disorder, the study of SOC is also a significant subject. SOC is known to play an important role in both theoretical and experimental investigations \cite{49,50,51,52}. For example, SOC can tune the bandgap and Fermi velocity by adjusting external conditions \cite{49}. It can also induce strong variations in magnetoresistance \cite{50}, enable dissipationless transport \cite{51}, and even drive topological phase transitions \cite{52}. However, the influence of SOC on anomalous quantum oscillations remains largely unexplored. This has motivated a further theoretical study of how SOC affects anomalous quantum oscillations in InAs/GaSb quantum wells.

\begin{figure}[htbp]
    \centering
    \includegraphics[width=0.48\textwidth]{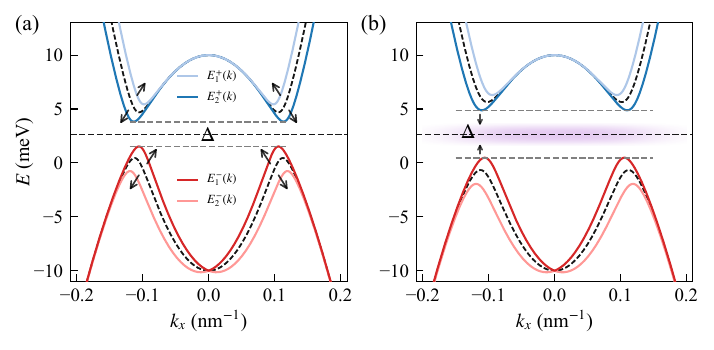}
    \caption{(Color online) Band structure of InAs/GaSb quantum wells with a direct bandgap $\Delta$. The black dashed curves represent the bands without SOC, and the red and blue solid curves correspond to the spin-split bands with SOC. (a) Clean case. Black arrows indicate the direction of SOC-induced band splitting. (b) Disordered case. The purple-shaded region schematically indicates the presence of disorder. The chemical potential $\mu$ is denoted by the horizontal black dashed line.}
    \label{fig:1}
\end{figure}

In this work, based on an effective four-band model, we investigate the effects of SOC on anomalous quantum oscillations in clean and disordered InAs/GaSb quantum wells (Fig.~\ref{fig:1}). The specific contribution of SOC in InAs/GaSb quantum wells is analytically computed based on perturbation theory. By comparing the evolution of the DOS in the absence and presence of SOC, the origin of anomalous quantum oscillations in the clean limit of the InAs/GaSb system is revealed. The specific role of SOC is then clarified in detail. Furthermore, in the disordered case, we investigate the anomalous quantum oscillations induced by disorder through analyses of the LL spectrum and the spectral function distribution. These results reveal how SOC modulates the quantum oscillations. Finally, under the coexistence of SOC and disorder, the parameter regime that enables optimal observation of oscillations is identified.

The rest of this paper is organized as follows: In Section \ref{sec:level2}, we introduce the model Hamiltonian and methods. Section \ref{sec:level3} demonstrates the modulation of anomalous quantum oscillations by SOC in both clean and disordered cases, and clarifies the underlying physical mechanisms. Finally, the main conclusions are summarized in Section \ref{sec:level4}.

\section{\label{sec:level2}Theoretical Framework}

We adopt the Bernevig-Hughes-Zhang model \cite{26,30,40}, which captures the essential properties of the InAs/GaSb quantum well system. The fundamental Hamiltonian $H_0(\boldsymbol{k})$ describes the orbital hybridization between the electron-like ($|e\rangle$) and hole-like ($|h\rangle$) subbands. The basis is given by $\{|e{\uparrow}\rangle, |h{\uparrow}\rangle, |e{\downarrow}\rangle, |h{\downarrow}\rangle\}$. The spin indices $\uparrow$ and $\downarrow$ denote spin-up and spin-down states, respectively. In the absence of a magnetic field, the Hamiltonian takes the form:
\begin{equation}
    \label{eq1}
    H_0(\boldsymbol{k}) =
    \begin{pmatrix}
        M_{1}(\boldsymbol{k}) & \beta k_{+}     &    0 & 0 \\
        \beta k_{-}        &  M_{2}(\boldsymbol{k})  & 0    & 0 \\
        0 &   0     &     M^{\ast}_{1}(-\boldsymbol{k}) & -\beta^{\ast} k_{-}  \\
        0 & 0  &   -\beta^{\ast} k_{+}     &  M^{\ast}_{2}(-\boldsymbol{k}) 
    \end{pmatrix},
\end{equation}
where the parabolic dispersions are $M_{1}(\boldsymbol{k}) = M_0 + \mu_{+} k^2$ and $M_{2}(\boldsymbol{k}) = -M_0 - \mu_{-} k^2$. Here, $ k^{2}=k_{x}^{2}+k_{y}^{2}$, $k_{\pm}=k_{x}\pm ik_{y}$, $\mu_\pm = \mu_0 \pm \delta\mu$, and $\beta$ characterizes the intersubband hybridization in $H_{0}$. 

In InAs/GaSb quantum wells, spatial separation between the electron-rich (InAs) and hole-rich (GaSb) layers breaks structural inversion symmetry. As a result, the structural inversion asymmetry (SIA) term becomes dominant, modifying the band structure and giving rise to Rashba spin-orbit coupling, described by \cite{26}:
\begin{equation}
    \label{eq2}
    H_{\text{SIA}}(\boldsymbol{k}) = 
    \begin{pmatrix}
        0 & 0  & -i\alpha k_{-} & 0   \\
        0 & 0  & 0  & 0 \\
        i\alpha k_{+} & 0 & 0 & 0  \\
        0 & 0  & 0  & 0 
    \end{pmatrix},
\end{equation}
where $\alpha$ represents the Rashba spin-orbit coupling strength.

To provide a complete understanding of the influence of SOC on anomalous quantum oscillations, we consider the InAs/GaSb system in the presence of disorder. Disorder is incorporated via a quasiparticle self-energy $\Sigma$ using the self-consistent Born approximation~\cite{14,45}, represented as a $4 \times 4$ matrix (details are given in Appendix~\ref{selfenergy}). The real part of $\Sigma(E)$ renormalizes the chemical potential and the inverted gap at $k = 0$, and for convenience, it is absorbed into $H_0$ in the following. The imaginary part forms a nonzero diagonal matrix:
\begin{equation}
    \label{eq3}
    \Sigma \approx
    \begin{pmatrix}
        -i\Gamma_e & 0 & 0 & 0 \\
        0 & -i\Gamma_h & 0 & 0 \\
        0 & 0 & -i\Gamma_e & 0 \\
        0 & 0 & 0 & -i\Gamma_h
    \end{pmatrix},
\end{equation}
where the diagonal terms $\Gamma_e$ and $\Gamma_h$ denote the disorder-induced scattering rates. In general, $\Gamma_e \neq \Gamma_h$ due to the distinct effective masses and scattering potentials of the two degenerate subbands. The total Hamiltonian, including both disorder and SOC, becomes non-Hermitian and can be written as:
\begin{equation}
\label{eq4}
H(\boldsymbol{k}) = H_0(\boldsymbol{k}) + H_{\text{SIA}}(\boldsymbol{k}) + \Sigma.
\end{equation}

To analyze the role of disorder, we first neglect $H_{\text{SIA}}$, which allows an exact analytical solution for the eigenenergies:
\begin{equation}
\begin{aligned}
\label{eq5}
E^{\pm}(\boldsymbol{k}) &=  \frac{1}{2} \left[ M_{1} + M_{2} - i (\Gamma_e + \Gamma_h)  \right] \\
&\pm \frac{1}{2} \sqrt{ \left( M_{1} - M_{2} - i (\Gamma_e - \Gamma_h) \right)^2 + 4 \beta^2 k^2 }.
\end{aligned}
\end{equation}
From Eq.~\eqref{eq5}, it is evident that the disorder-induced self-energy $\Sigma$ leads to level broadening and a finite quasiparticle lifetime through the imaginary components $-i(\Gamma_e \pm \Gamma_h)$. These terms significantly modify the spectral function and oscillatory behavior relative to the clean limit.

When SOC ($H_{\text{SIA}}$) is introduced as a perturbation, the energy levels acquire first-order corrections (see Appendix~\ref{SOC} for derivation):
\begin{equation}
\begin{aligned}
    E^{+}_{p}(\bm{k}) = E^{+}(\bm{k})& \pm \alpha k \mathcal{F}_{+}(k), \quad E^{-}_{p}(\bm{k}) = E^{-}(\bm{k}) \pm \alpha k \mathcal{F}_{-}(k),  \\
    &\mathcal{F}_{\pm}(k) = \frac{4\beta^{2}k^{2}}{4\beta^{2}k^{2}+(\delta_{\Gamma} \mp D_{\Gamma})^{2}}.
\end{aligned}
\label{eq6}
\end{equation}
Here, $\delta_{\Gamma} = M_1 - M_2 - i(\Gamma_e - \Gamma_h)$ and $D_{\Gamma} = \sqrt{\delta_{\Gamma}^2 + 4\beta^2 k^2}$. As illustrated in Fig.~\ref{fig:1}, spin degeneracy is lifted, and the band structure is reconstructed by the $\alpha k$-dependent terms, leading to a sequence of energy levels labeled as $E_1^+$, $E_2^+$, $E_1^-$, and $E_2^-$. The imaginary components $i(\Gamma_h - \Gamma_e)$ further modulate the distribution of in-gap states, demonstrating that SOC plays a dual role in both band reconstruction and in-gap state tuning.

In the following, we employ the approach of Knolle and Cooper \cite{30} to calculate the energy levels in the presence of a perpendicular magnetic field. The magnetic field $B$ is incorporated via the vector potential $\mathbf{A}$ using the minimal coupling scheme, in which the crystal momentum $\hbar \mathbf{k}$ is replaced by $\boldsymbol{\Pi} = \hbar \mathbf{k} + \tfrac{e}{c}\mathbf{A}$. The momentum operators are then expressed in terms of the ladder operators: $k_+ \rightarrow \frac{\sqrt{2}}{l_B} \hat{a}^\dagger$, $k_- \rightarrow \frac{\sqrt{2}}{l_B} \hat{a}$ and $k^2 \rightarrow \frac{2}{l_B^2} \left(\hat{a}^\dagger \hat{a} + \frac{1}{2}\right)$, where the magnetic length is defined as $l_B = \sqrt{\hbar c / e|B|} \approx 26~\text{nm}/\sqrt{B(\mathrm{T})}$. We adopt an ansatz wave function $|\Psi_n\rangle = \left( a_n|n\rangle,\ b_n|n-1\rangle,\ c_n|n+1\rangle,\ d_n|n+2\rangle \right)^T$, constructed from harmonic oscillator basis states $|n\rangle$. Using this ansatz, the original Hamiltonian $H(\bm{k})$ is transformed into the magnetic Hamiltonian $H_n$ for a given magnetic field $B$. Disorder effects are also incorporated into this Hamiltonian. The explicit form of $H_n$ is provided in the Appendix~\ref{magneticHamiltonian}. The energy levels $E_n$ are then obtained by solving the eigenvalue equation $\hat{H}_n|\Psi_n\rangle = E_n|\Psi_n\rangle$, resulting in the Landau level spectrum as a function of the index n.

Both dHvA and SdH oscillations can be characterized through the oscillatory behavior of the low-energy density of states (LEDOS) \cite{13,15}. The corresponding expressions are:
\begin{equation}
    \label{eq7}
    D(T) = - \int_{-\infty}^{\infty} d\omega\, \frac{\partial n_{\mathrm{F}}(\omega-\mu, T)}{\partial \omega} A(\omega),
\end{equation}
where $n_{\mathrm{F}}(\omega-\mu, T) = \left[ e^{(\omega-\mu)/(k_{B} T)} + 1 \right]^{-1}$. The spectral function is given by
\begin{equation}
    \label{eq8}
    A(\omega) = -\frac{N_B}{\pi}  \mathrm{Im} \sum_{n,b} \frac{1}{\omega - E_{n}^{b}},
\end{equation}
where $E^{b}_{n}$ denoting the eigenenergies obtained from numerical diagonalization of $H_{n}$, $b$ labeling the eigenvalues of each $H_{n}$, and $N_B = Be/h$ the degeneracy per Landau level.

Simulating the dHvA or SdH oscillations observed in experiments requires taking into account the material scattering mechanisms and electron interactions, which are beyond the scope of our model. Instead, we compute the oscillations of the LEDOS as a robust proxy for these oscillations. LEDOS is related to various physical quantities at finite temperature, such as the magnetic susceptibility and the resistivity. Thus, LEDOS oscillations necessarily induce corresponding oscillations in these quantities.

\section{\label{sec:level3} NUMERICAL RESULTS}
In this section, the LL spectra, LEDOS oscillations, and spectral function are studied numerically. In experiments, the InAs/GaSb quantum well exhibits an inverted band structure with a heavy-hole band of larger effective mass. To capture these features, we set $M_{0} = -10~\mathrm{meV}$, $\mu_0 = 819~\mathrm{meV\cdot nm}^2$, and $\delta\mu = 216~\mathrm{meV\cdot nm}^2$~\cite{30,40}, with the chemical potential fixed at $\mu = 2.644~\mathrm{meV}$. 

\subsection{Anomalous Quantum Oscillations in the Absence of SOC}
\begin{figure}[htbp]
    \centering
    \includegraphics[width=0.48\textwidth]{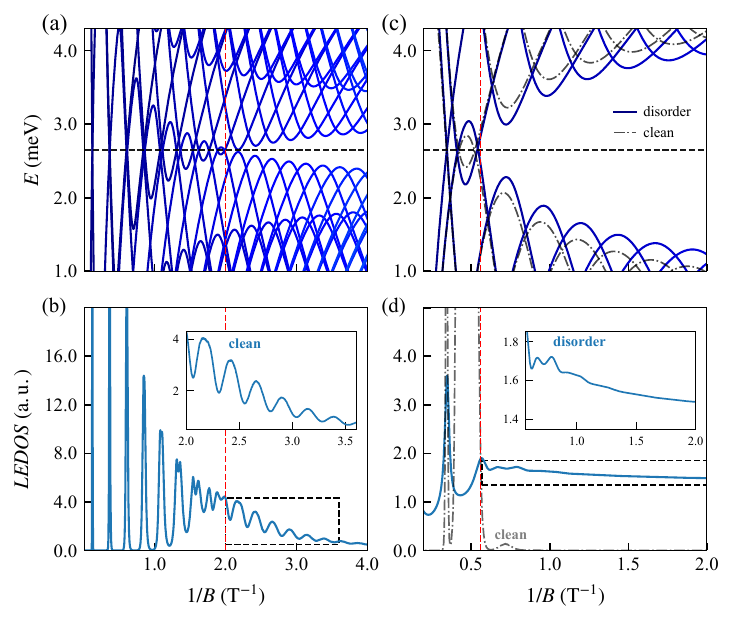}
    \caption{(Color online) LL spectra and LEDOS in the absence of SOC. (a,b) Clean case ($\beta = 5$ meV$\cdot$nm) at temperature $T = 0.3\Delta$. (c,d) Clean (gray dash-dotted curves) and Disordered (blue solid curves) cases at $\beta = 20$ meV$\cdot$nm, calculated at an extremely low temperature $T = 0.003\Delta$ to emphasize disorder effects. The red dashed line marks the critical magnetic field $B_c$ separating the metallic and insulating regimes. The insets in (b) and (d) highlight the emergence of anomalous quantum oscillations in the insulating regime.}
    \label{fig:2}
\end{figure}

Figure~\ref{fig:2} presents the LL spectra and the corresponding LEDOS oscillations for both the clean and disordered cases in the absence of SOC. The red dashed line represents $1/B_c$, the inverse of the critical magnetic field $B_c$.
For the InAs/GaSb system in the clean limit, the LL spectrum and the corresponding LEDOS oscillation are plotted in Figs.~\ref{fig:2}(a) and \ref{fig:2}(b).
In Fig.~\ref{fig:2}(a), the bandgap is closed beyond a critical field $B_c$, where $1/B_c = 2.0$~T$^{-1}$.
For $1/B < 1/B_c$, the strong magnetic field results in a large spacing between LLs. Compared to this large spacing, the hybridization potential ($\beta k$) is too small to open the bandgap. Thus, the spectrum near the chemical potential remains metallic. As the magnetic field decreases ($1/B > 1/B_c$), level repulsion overcomes the small LL spacing, opening a finite hybridization gap and driving the system into an insulating regime. 
Therefore, it is clearly observed that the magnetic field induces a metal–insulator transition.

The corresponding LEDOS oscillation is shown in Fig.~\ref{fig:2}(b). 
For $1/B < 1/B_c$, the LEDOS exhibits well-defined oscillations, consistent with the characteristics of conventional quantum oscillations. 
When $1/B > 1/B_c$, such oscillations are expected to vanish because no LLs cross the chemical potential. 
Nevertheless, clear oscillations can still be observed in Fig.~\ref{fig:2}(b), revealing the emergence of quantum oscillations in the insulating regime, named anomalous quantum oscillations \cite{13}.

To see this more clearly, the anomalous quantum oscillations are plotted in the inset of Fig.~\ref{fig:2}(b), which exhibits six sharp, well-defined peaks with a maximum value of 4.0. The average oscillation amplitude, calculated by the mean difference between adjacent peaks and valleys, is approximately 1.65. 
These results demonstrate that the anomalous oscillations are pronounced and regular. 

Next, we increase the hybridization strength to $\beta = 20$. 
The LL spectrum and the corresponding LEDOS oscillation are plotted in Figs.~\ref{fig:2}(c) and \ref{fig:2}(d), with gray dash-dotted curves. 
In Fig.~\ref{fig:2}(c), with the enlarged bandgap, the critical field moves from $1/B_c = 2.0$~T$^{-1}$ to $0.58$~T$^{-1}$. 
The corresponding LEDOS is shown in Fig.~\ref{fig:2}(d). Conventional LEDOS oscillations appear in the metallic regime, whereas anomalous quantum oscillations vanish when the bandgap becomes sufficiently large.

Based on this observation, we infer that, in the clean limit, the emergence of anomalous quantum oscillations requires a narrow bandgap, consistent with the observation reported by Zhang et al.~\cite{13}. 
In this case ($\beta = 5$), the small bandgap is on the order of $k_B T$, which leads to thermally excited states at the chemical potential~\cite{30}. 
Under magnetic fields, the periodically narrowing hybridization gap drives these thermally activated states to oscillate, thereby leading to the observation of anomalous quantum oscillations in the insulating regime. 
This understanding naturally raises the question of whether other realistic factors, such as disorder, can induce anomalous oscillations when the bandgap becomes large.

To address this question, we investigate the effect of disorder in the system ($\beta=20$). 
Following the approach used in heavy-fermion systems~\cite{14}, we incorporate disorder through self-energy $\Sigma \approx \mathrm{diag}(-i\Gamma_e,-i\Gamma_h,-i\Gamma_e,-i\Gamma_h)^T$ and study its impact on the LEDOS oscillations in the InAs/GaSb system.
Due to the large effective mass of holes, the hole scattering rate is high \cite{35}.
Accordingly, $\Gamma_e$ is fixed at $0.1\Delta$, and the strength of the disorder potential is quantified through $\Gamma_h$, which is taken as $0.8\Delta$ for the present calculation.

For $\beta = 5$, the narrow bandgap makes the system highly susceptible to disorder, easily driving it to a metallic phase. 
By comparison, for $\beta = 20$, the bandgap remains large even with moderate disorder strength, so the system retains an insulating regime. 
Therefore, we introduce disorder effects into the case of $\beta = 20$ to examine whether it can induce anomalous quantum oscillations.

When disorder is introduced, the LL spectrum is plotted in Fig.~\ref{fig:2}(c) with blue solid lines. 
It is observed that the disorder does not alter the critical field, i.e., $1/B_c = 0.58~\mathrm{T}^{-1}$. However, it broadens the tails of the LLs, shifting the LLs toward the chemical potential. 
The LEDOS oscillations are plotted in Fig.~\ref{fig:2}(d) with blue solid lines.
For $1/B < 1/B_c$, it exhibits conventional oscillations, while for $1/B > 1/B_c$, disorder induces a finite LEDOS, indicating the presence of in-gap states. 
Interestingly, weak oscillations also appear in the insulating regime. These anomalous oscillations are visible in the inset of Fig.~\ref{fig:2}(d), which shows two weak and irregular peaks with a maximum value of approximately 1.70.
Quantitatively, the average amplitude is approximately 0.05. 
This signals the emergence of weak anomalous quantum oscillations in the disordered case.

\subsection{Anomalous Quantum Oscillations in the Presence of SOC}
\begin{figure}[htbp]
    \centering
    \includegraphics[width=0.48\textwidth]{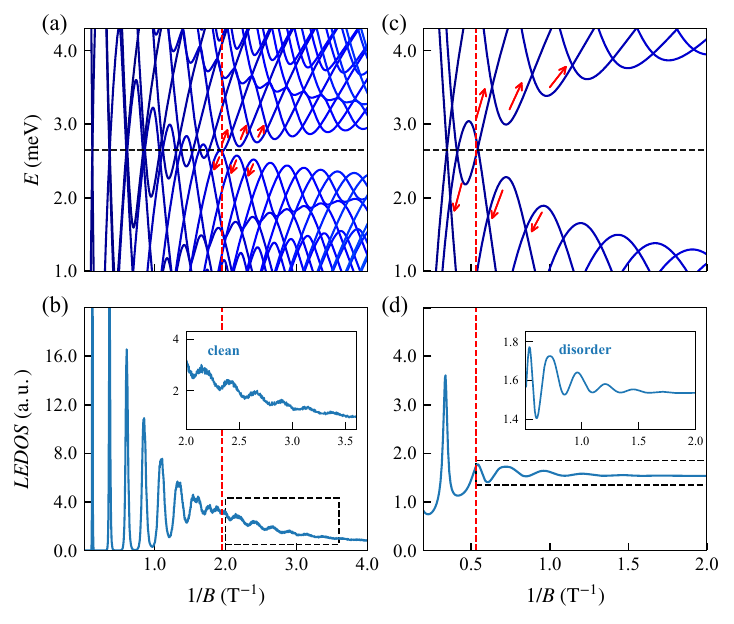}
    \caption{(Color online) LL spectra and LEDOS in the presence of SOC. (a,b) Clean case with $\beta=5$: SOC ($\alpha=4$ meV$\cdot$nm) induces bandgap expansion, suppressing the anomalous LEDOS oscillations. (c,d) Disordered case with $\beta=20$: SOC ($\alpha=16$ meV$\cdot$nm) enlarges the bandgap, while enhancing anomalous LEDOS oscillations compared to the case without SOC. Red arrows indicate the direction of bandgap widening. The inset highlights the anomalous quantum oscillations in the insulating regime.}
    \label{fig:3}
\end{figure}

We have demonstrated that anomalous quantum oscillations emerge in both clean and disordered InAs/GaSb quantum wells. 
We now turn to the influence of SOC, which plays a key role in InAs/GaSb systems. 
In particular, it is of great interest to understand how SOC affects anomalous quantum oscillations in these two distinct cases.
Figure~\ref{fig:3} shows the effects of SOC on the LL spectra and LEDOS for both the clean and disordered InAs/GaSb quantum wells. 
Figures~\ref{fig:3}(a) and~\ref{fig:3}(b) correspond to the clean cases, and Figures~\ref{fig:3}(c) and~\ref{fig:3}(d) correspond to the disordered cases. 

In the clean limit, upon the introduction of SOC,  the LLs are shifted along the directions indicated by the red arrows, as shown in Fig.~\ref{fig:3}(a). The shifts of LLs stem from the change of the energy band by SOC.
This shift causes the original critical field, $1/B_c$, to move from $2.0$~T$^{-1}$ to $1.9$~T$^{-1}$. 
For $1/B < 1/B_c$, the LL spectrum is almost the same as that of the system without SOC, resulting in similar LEDOS oscillations. 
In contrast, for $1/B > 1/B_c$, the shift of the LLs induced by SOC leads to a widening of the bandgap. This bandgap enlargement causes noticeable changes in the anomalous quantum oscillations.

Specifically, only five peaks are visible in the inset of Fig.~\ref{fig:3}(b), compared to six in the inset of Fig.~\ref{fig:2}(b). The maximum peak also drops to approximately 3.0, accompanied by pronounced phase smearing. Moreover, the average amplitude decreases from approximately 1.65 to 0.65.
These changes originate from the significant modification of the Landau-level band structure by SOC.
The bandgap is enlarged by SOC, exceeding the order of $k_B T$.
As a result, the number of thermally excited states is reduced at the chemical potential, thereby suppressing the anomalous LEDOS oscillations.
In other words, the anomalous quantum oscillations are suppressed by SOC in this case.

Next, we focus on the disordered case.
In Fig.~\ref{fig:3}(c), the introduction of SOC induces pronounced shifts of the LLs, as indicated by the red arrows. 
This also causes the original critical $1/B_c$ to shift from $0.58$~T$^{-1}$ to $0.53$~T$^{-1}$. 
For $1/B < 1/B_c$, the overlap of LLs near the chemical potential is reduced, but the LEDOS oscillations in the metallic regime remain nearly unchanged.

For $1/B > 1/B_c$, the introduction of SOC leads to an obvious change for the anomalous LEDOS oscillations, as shown in Fig.~\ref{fig:3}(d). Specifically, compared with the case without SOC, the anomalous LEDOS oscillations change noticeably upon introducing SOC: the number of well-defined peaks increases from two to four, the maximum peak rises from about 1.7 to 1.8, and the average amplitude increases from roughly 0.05 to nearly 0.25, as shown in the inset of Fig.~\ref{fig:3}(d). Strikingly, this demonstrates a strong enhancement of anomalous oscillations induced by SOC in the disordered case.

The different behavior between the clean and disordered cases reveals the opposing effects of SOC on anomalous quantum oscillations. 
In the clean limit, SOC primarily widens the bandgap, thereby reducing the number of thermally activated states at the chemical potential. 
In the disordered case, after introducing SOC, even though the bandgap is further enlarged, the anomalous quantum oscillations are enhanced markedly.
This naturally raises an important question: how does SOC enhance anomalous quantum oscillations in the presence of disorder?

\subsection{Reason for the Enhancement of Anomalous Oscillations by SOC in the Disordered Case}
\begin{figure}[htbp]
    \centering
    \includegraphics[width=0.48\textwidth]{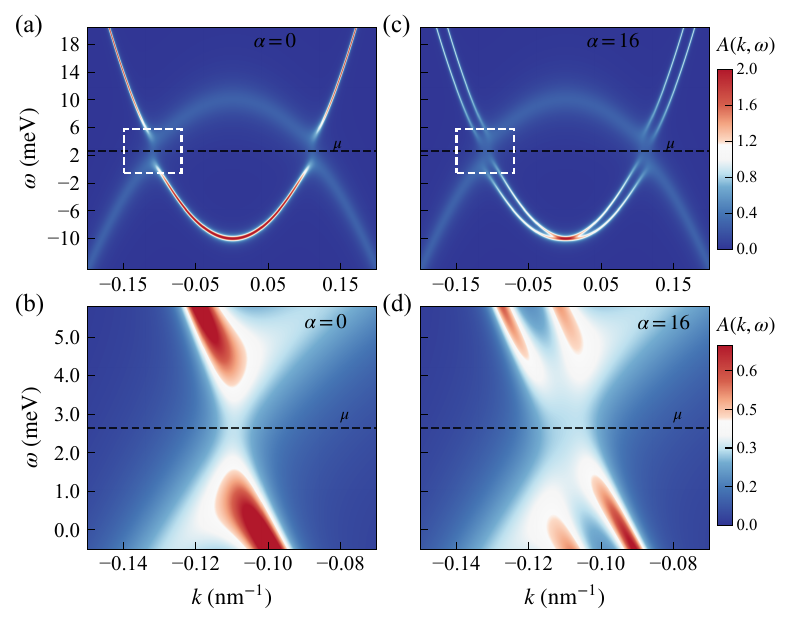}
    \caption{(Color online) Spectral functions of the disordered case with $\Gamma_h = 0.8 \Delta$ in the absence of a magnetic field. (a) Without SOC ($\alpha=0$) and (c) with SOC ($\alpha=16$~meV$\cdot$nm), showing the overall spectral features. (b,d) Enlarged views of the white dashed rectangles in (a) and (c), respectively, display detailed gap structures.}
    \label{fig:4}
\end{figure}
To understand the enhancement of anomalous oscillations induced by SOC, we first analyze the spectral function $A(\mathbf{k}, \omega) = -\frac{1}{\pi} \operatorname{Im} \left[ 1 / (\omega - H(\mathbf{k}) - \Sigma) \right]$. 
The spectral functions of the disordered case are shown in Fig.~\ref{fig:4}. 
In Fig.~\ref{fig:4}(a), without SOC, two parabolic spectral regions are clearly observed. 
Details of the hybridization region are shown in Fig.~\ref{fig:4}(b). 
When SOC is included, a clear splitting is observed, which separates the previously degenerate spectral regions into two distinct branches, as shown in Fig.~\ref{fig:4}(c).
This splitting is obvious in the magnified view of Fig.~\ref{fig:4}(d), where the SOC-induced splitting broadens the states near the bandgap, enhancing the spectral weight near the chemical potential. 
The broadened states are highlighted by the bright white regions.
This enhancement can also be understood from the SOC-modified term $\alpha k \mathcal{F}_{\pm}(k)$ in Eq.~\eqref{eq6}. The imaginary part of $\mathcal{F}{\pm}(k)$ can be expressed as:
\begin{equation}
\mathrm{Im}(\mathcal{F}_{\pm}(k))\approx \pm \frac{2 \Delta_{\Gamma}(4\beta^{2}k^{2})(\delta \mp D)}{D[4\beta^{2}k^{2} + (\delta \mp D)^{2}]^{2}},
\label{eq9}
\end{equation}
where $\Delta_{\Gamma} = \Gamma_h - \Gamma_e$, $\delta= M_{1} - M_{2}$ and $D = \sqrt{\delta^{2} + 4\beta^{2}k^{2}}$. 
The real part of the SOC-modified term $\alpha k \mathcal{F}_{\pm}(k)$ in Eq.~\eqref{eq6} modifies the band structure, leading to a splitting of previously degenerate spectral regions into two subbands. 
The imaginary part of $\mathcal{F}_{\pm}(k)$ (Eq.~\eqref{eq9}) determines the broadening size of these subbands, influencing the increase in spectral weight near the chemical potential within the gap. 
This enhanced spectral weight can potentially lead to more pronounced anomalous oscillations compared with the case without SOC.

\begin{figure}[htbp]
    \centering
    \includegraphics[width=0.48\textwidth]{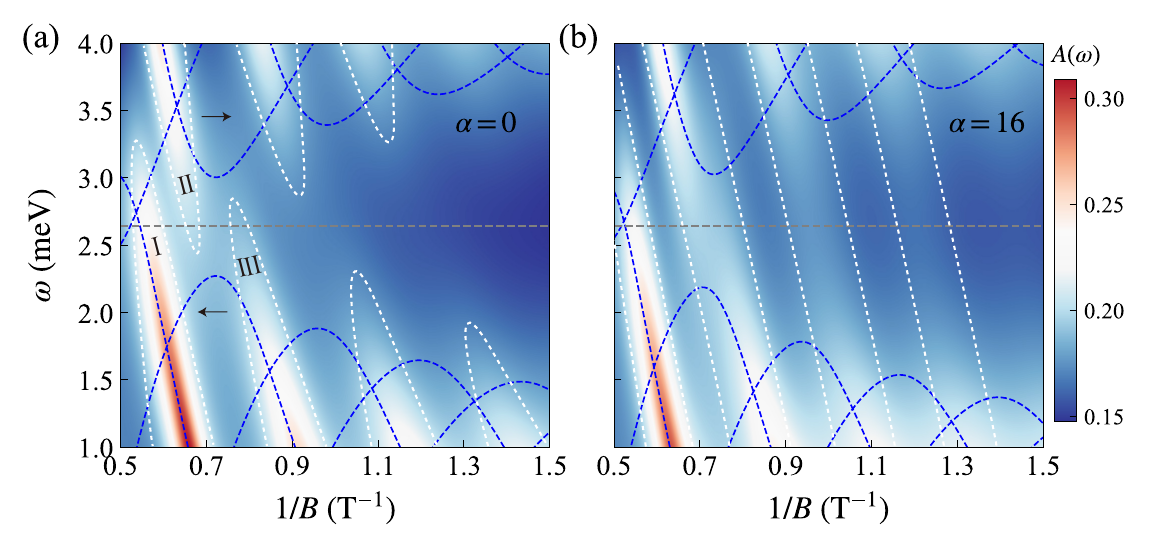}
    \caption{(Color online) Spectral functions with $\Gamma_h = 0.8 \Delta$ in the presence of magnetic field: (a) without SOC and (b) with SOC. The blue dashed curves indicate Landau levels; the white dashed contours outline the high-density regions.}
    \label{fig:5}
\end{figure}

In the following, we focus on the influence of SOC on the in-gap states in the presence of magnetic field. With the magnetic field, the spectral function is computed by Eq.~\eqref{eq8}.
For the disordered case, the LLs and the distributions of the spectral function are shown without and with SOC in Figs.~\ref{fig:5}(a) and \ref{fig:5}(b), respectively.

The hole scattering rate is set to a large value ($\Gamma_h = 0.8\Delta$), which results in the spectral weight being predominantly concentrated near the hole-like LLs, as outlined by the white dashed envelope in Fig.~\ref{fig:5}.
It is clearly seen that several high-intensity regions emerge around or even cross the chemical potential, forming an interlaced pattern in Fig.~\ref{fig:5}(a). The first three regions of high spectral weight from left to right are labeled I, II, and III. We observe that region I exhibits the strongest spectral intensity, reflecting a large value of $A(\omega)$. For Region~I, the magnetic field reaches $1.8$~T, resulting in a large Landau-level degeneracy $N_B$, which directly enhances $A(\omega)$. As the magnetic field decreases, the spectral weight of regions II and III gradually decreases.

There are only the first three regions of high-intensity intersecting with the chemical potential [Fig.~\ref{fig:5}(a)]. According to Eq.~\eqref{eq7}, regions with high spectral weight naturally produce local peaks in the LEDOS spectrum, whereas regions with low spectral weight correspond to valleys. We note that the three high-intensity regions crossing the chemical potential correspond to the three visible peaks observed in the LEDOS inset of Fig.~\ref{fig:2}(d), while the non-crossing regions contribute almost nothing to the oscillatory signal. Thus, this spectral-weight pattern explains why only weak but discernible oscillations appear in the LEDOS spectrum.

Interestingly, when the LLs are shifted by SOC, the spectral-weight regions are also moved, as indicated by the black arrows in Fig.~\ref{fig:5}(a). Under the effect of SOC, the spectral pattern undergoes a qualitative change in Fig.~\ref{fig:5}(b). Strikingly, the previously interlaced spectral-weight pattern [Fig.~\ref{fig:5}(a)] evolves into a highly ordered configuration [Fig.~\ref{fig:5}(b)], manifested as alternating bright and dark stripes across the gap region. For example, the originally interlaced regions I and II merge into a single continuous region. This pronounced redistribution sharply enhances the difference between regions of high and low spectral density, thereby enhancing the LEDOS oscillation amplitude. In particular, the oscillatory peaks and valleys in Fig.~\ref{fig:3}(d) perfectly match the regions of enhanced and reduced spectral weight in Fig.~\ref{fig:5}(b). Such an obvious correspondence reveals that in the disordered case, the SOC-induced redistribution of in-gap states is the microscopic origin of the enhanced anomalous quantum oscillations.

To sum up, the opposite behaviors of anomalous quantum oscillations observed in Figs.~\ref{fig:3}(c) and \ref{fig:3}(d) arise naturally from the distinct dominant roles that SOC plays. In the clean limit, a narrow bandgap is the key factor inducing anomalous quantum oscillations, and the introduction of SOC widens the bandgap, thereby suppressing these oscillations. In the disordered case, anomalous oscillations are driven by disorder-induced in-gap states, and SOC causes these states to become more densely distributed in certain regions than in its absence, thereby enhancing the anomalous quantum oscillations.

\subsection{Anomalous Quantum Oscillations at Different Temperatures}

\begin{figure}[htbp]
    \centering
    \includegraphics[width=0.40\textwidth]{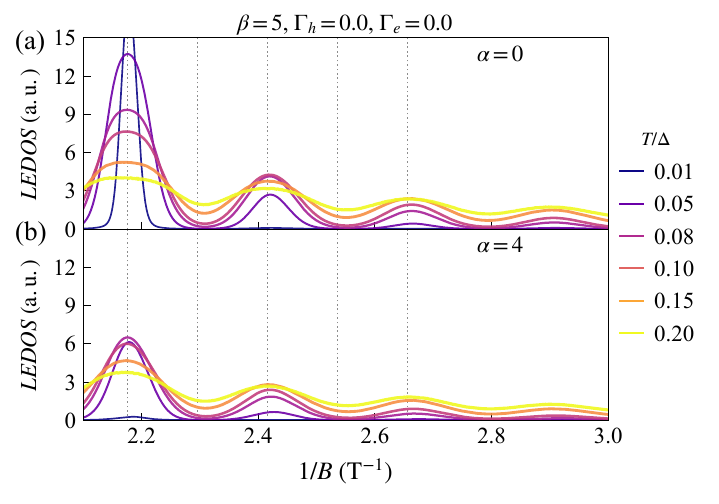}
    \vspace{0.0cm}
    \includegraphics[width=0.40\textwidth]{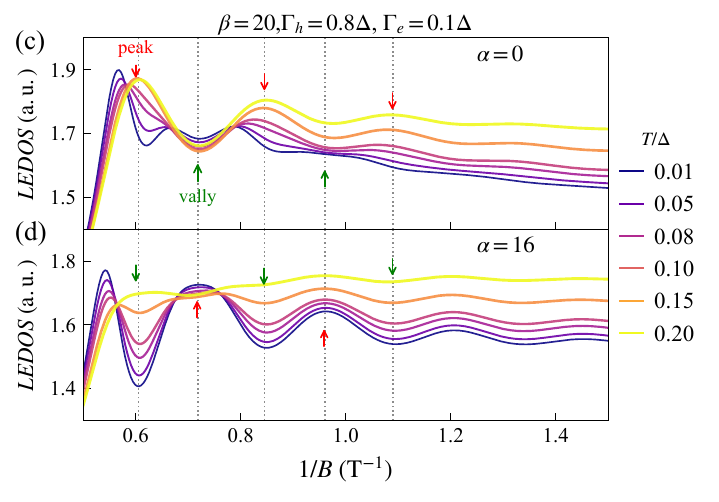}
    \caption{(Color online)  Temperature dependence of the quantum oscillations in insulating regimes. (a,b) Clean case: (a) without SOC and (b) with SOC, showing amplitude suppression. (c,d) Disordered case: (c) without SOC and (d) with SOC, where SOC enhances the oscillations at low temperatures. The red and green arrows represent the oscillation peaks and valleys, respectively.}
    \label{fig:6}
\end{figure}

To investigate the stability of the SOC effect on anomalous quantum oscillations, we examine the temperature dependence of the oscillations in both the clean and disordered cases, as shown in Fig.~\ref{fig:6}. In standard Lifshitz-Kosevich theory \cite{04,12}, the effect of a finite temperature is equivalent to phase smearing, which reduces the oscillation amplitude by a thermal reduction factor $R_T$. This factor is given by
\begin{equation}
    \label{10}
    R_T = \frac{\lambda T / B}{\sinh(\lambda T / B)}, \quad \lambda = \frac{2 \pi^2 k_B m^*}{\hbar e},
\end{equation}
indicating that the oscillation amplitude decreases monotonically with increasing temperature.

For the clean case, the anomalous quantum oscillations without and with SOC are presented in Figs.~\ref{fig:6}(a) and \ref{fig:6}(b), respectively. As shown in Fig.~\ref{fig:6}(a), at low temperatures ($T \leq 0.1\Delta$), the anomalous oscillations are pronounced, with the maximum amplitude exceeding 15 at $T = 0.01\Delta$. As the temperature increases, the oscillation amplitude gradually decreases due to phase smearing, which is described by the thermal reduction factor $R_T$.
When SOC is introduced, the previously strong anomalous oscillations are obviously suppressed, as shown in Fig.~\ref{fig:6}(b), consistent with our earlier observation that SOC suppresses anomalous quantum oscillations in the clean limit.

For the disordered case, the anomalous quantum oscillations without and with SOC are presented in Figs.~\ref{fig:6}(c) and \ref{fig:6}(d), respectively. In Fig.~\ref{fig:6}(c), for the four curves with $T \leq 0.1 \Delta$, each curve exhibits only a single oscillation peak, and all of these peaks are located near $1/B = 0.56~\mathrm{T}^{-1}$. The amplitude of these oscillations decreases gradually as the temperature increases. Interestingly, at higher temperatures [$T = (0.15, 0.20) \Delta$], the anomalous oscillations become more pronounced than those in the previous low-temperature cases. This behavior is similar to the thermally activated oscillations observed in the clean limit. The introduction of disorder reduces the bandgap, and with $T$ increasing enough, $k_B T$ can roughly match the gap, giving rise to the observed oscillations.

When SOC is introduced, the anomalous quantum oscillations at $T \leq 0.1 \Delta$ are significantly enhanced in Fig.~\ref{fig:6}(d), consistent with our previous observation in the disordered case. As the temperature increases, the oscillation amplitude gradually decreases. When $T$ is beyond $0.15 \Delta$, the anomalous oscillations almost disappear, and the LEDOS as a function of $1/B$ tends to vary linearly at the same time. This amplitude reduction arises from the phase smearing effect. The analysis of these two cases indicates that for $T \leq 0.15 \Delta$, the effect of SOC on anomalous quantum oscillations is robust.

Moreover, by comparing Figs.~\ref{fig:6}(c) and \ref{fig:6}(d), we observe that the introduction of SOC induces a noticeable phase shift in the oscillations, as evidenced by the peak-to-valley inversion occurring at specific $1/B$ positions. The maximum phase shift can reach nearly $\pi$, indicating a significant modification of the phase. According to the Onsager relation \cite{003}, the observed peak-to-valley inversion reflects a change of the Berry phase. This $\pi$-phase shift likely involves a combined effect of temperature, SOC, and disorder, among other factors. We do not intend to explore this in detail here; it will be the focus of our future work.

\subsection{Temperature-Disorder Phase Diagram of Anomalous Quantum Oscillations}
\begin{figure}[htbp]
    \centering
    \includegraphics[width=0.40\textwidth]{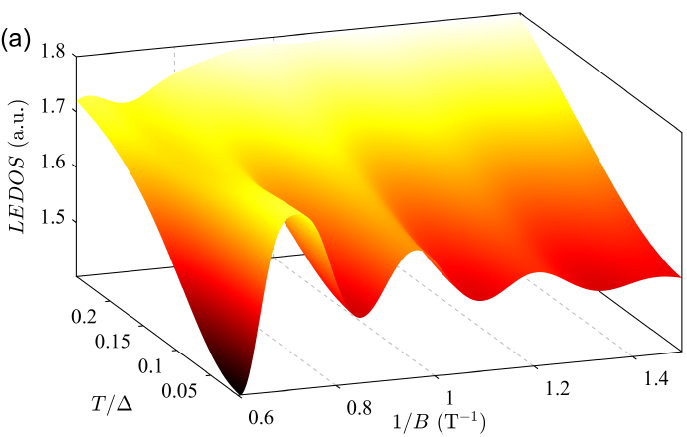}
    \vspace{0.0cm}
    \includegraphics[width=0.40\textwidth]{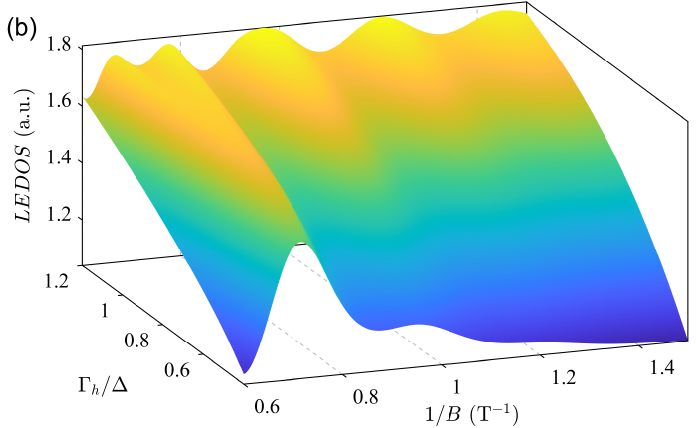}
    \vspace{0.0cm}
    \includegraphics[width=0.40\textwidth]{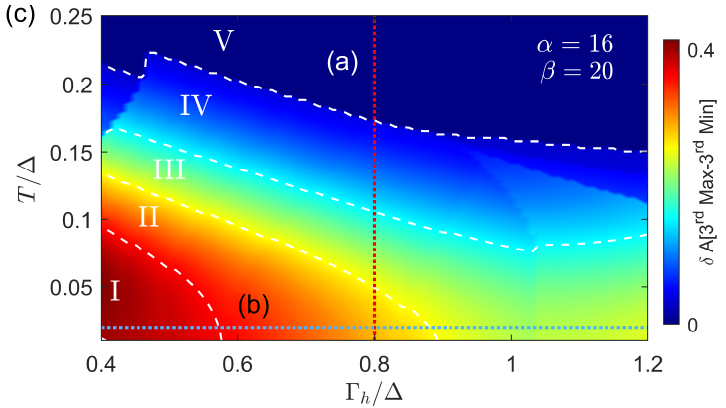}
    \caption{(Color online) (a) Temperature evolution of the LEDOS oscillations at a fixed disorder strength $\Gamma_h = 0.8\Delta$. (b) Evolution of the LEDOS as a function of disorder strength $\Gamma_h$ at a fixed temperature $T = 0.02\Delta$. (c) Phase diagram of $\delta A$ in the ($\Gamma_h, T$) parameter space, where the red dashed lines represent $\delta A$ extracted from the oscillation curves in Fig.~\ref{fig:7}(a), and the blue dashed lines represent $\delta A$ extracted from the oscillation curves in Fig.~\ref{fig:7}(b). Based on the variation of $\delta A$, the diagram is divided into five regions, labeled I-V.}
    \label{fig:7}
\end{figure}

Based on the previous analysis, we have verified the effect of SOC on anomalous quantum oscillations in different cases. Compared with the suppression of the anomalous oscillations by SOC in the clean limit, in the presence of disorder, the enhancement of oscillations by SOC is more attractive to us. Besides the strength of disorder, temperature is an essential factor in experiments. We show the anomalous quantum oscillations with SOC as a function of temperature and disorder strength, as presented in Fig.~\ref{fig:7}.

For $\Gamma_h = 0.8\Delta$, the temperature evolution of the LEDOS oscillations is calculated. The thermal reduction factor $R_T$ introduced in Eq.~\eqref{10} characterizes the temperature-induced phase smearing, which increases with temperature. In Fig.~\ref{fig:7}(a), pronounced oscillations are clearly seen at low temperatures ($T < 0.1\Delta$), where the $R_T$ is weak. In this regime, anomalous quantum oscillations are well captured. As the temperature increases, the smearing effect of $R_T$ becomes larger, leading to a gradual decay of the oscillation amplitude. Meanwhile, the increase in temperature results in a larger number of thermally broadened states within the bandgap, thereby increasing the overall LEDOS. When the temperature is further increased, the oscillations are almost completely suppressed around $T = 0.25\Delta$.

In addition, we also show the evolution of the LEDOS oscillations as a function of the disorder scattering rate at $T = 0.02\Delta$. In Fig.~\ref{fig:7}(b), within a weak disorder regime ($\Gamma_h = 0.2\Delta$-$0.5\Delta$), only a single pronounced anomalous oscillation peak is observed, with a relatively large amplitude. For this regime, the in-gap states near the chemical potential are sparse, so the oscillatory signal appears only in strong magnetic-field regions where the bandgap is narrow. As the disorder strength increases, additional in-gap states emerge, giving rise to extra oscillation peaks. When $\Gamma_h$ exceeds $1.05\Delta$, the bandgap begins to close, shifting the critical magnetic field $B_c$ toward lower values. Then, the oscillations observed at stronger magnetic fields correspond to conventional quantum oscillations rather than anomalous ones.

Finally, a phase diagram of the LEDOS in the $(\Gamma_h, T)$ parameter space is presented in Fig.~\ref{fig:7}(c). The main oscillation amplitude is quantified by $\delta A$, defined as the difference between consecutive maxima and minima of the LEDOS (details can be found in Appendix~\ref{deltaA}). Based on the evolution along the two axes of temperature and disorder strength, the phase diagram of $\delta A$ is divided into five regions, labeled I-V, in which the influences of temperature and disorder on quantum anomalous oscillations are clearly illustrated.

When $\Gamma_h \in [0.4\Delta, 0.6\Delta)$, the system evolves from Region~I to Region~V as the temperature increases. During this evolution, $\delta A$ gradually decreases. This demonstrates that temperature plays a strong suppressing role on the anomalous quantum oscillations. The underlying mechanism is the thermal damping factor $R_T$: increasing temperature enhances the thermal smearing effect, which reduces the oscillation amplitude. 
After adjusting  $\Gamma_h \in [0.6\Delta, 0.9\Delta)$, Region~I disappears in this range. The system evolves from Region~II to Region~V as the temperature rises. Here, the suppression of oscillations is still dominated by thermal smearing, which gradually washes out the oscillatory features. 
For $\Gamma_h \in (0.9\Delta, 1.2\Delta]$, only Regions~III-V are present in this range. This indicates that the amplitudes of oscillation decrease with stronger disorder. As the temperature continues to increase, the weakened oscillations are further suppressed and eventually vanish. Overall, both temperature and disorder significantly suppress the anomalous quantum oscillations. Temperature acts through thermal broadening, while disorder acts through spectral broadening. The phase diagram clearly illustrates their combined damping effect.

It is worth emphasizing that Region~I is not the optimal parameter space for observing well-defined oscillations in experiments. Although the value of $\delta A$ is the largest in this region, the large amplitude originates from a single pronounced oscillation peak with a large peak-to-valley difference [see Fig.~\ref{fig:7}(b)]. Because of the lack of useful periodic information, a single-peak structure is unsuitable for analyzing quantum oscillations.

In comparison, Region~II represents the most favorable parameter regime. The values of $\delta A$ in this region are the second largest, and the oscillations exhibit a multi-peak structure, as seen in Figs.~\ref{fig:7}(a) and \ref{fig:7}(b). Notably, for $T < 0.1\Delta$, the oscillations display well-resolved, peak-rich features. These features are highly advantageous for extracting and analyzing quantum oscillations. Therefore, the anomalous quantum oscillations are most optimally observed in the parameter window of  
$T < 0.1\Delta, 0.6\Delta < \Gamma_h < 0.9\Delta$, which provides clear, well-resolved multi-peak oscillations. This result offers practical guidance for experimental detection.

%===============================================================   Sec5：conclusions  =========================================================================================
\section{\label{sec:level4}Conclusions}
In this work, we have investigated the effect of SOC on anomalous quantum oscillations in InAs/GaSb quantum wells, considering both clean and disordered cases. Our analysis reveals that SOC plays a dual role, modulating the bandgap and redistributing in-gap states. In the clean limit, SOC widens the bandgap, suppressing the anomalous quantum oscillations by reducing the contribution of thermally excited carriers. In the disordered case, SOC concentrates the in-gap spectral weight in certain regions, forming a pronounced alternating bright-and-dark pattern. This redistribution, in turn, enhances the anomalous LEDOS oscillations. 
In the disordered case, SOC induces a phase shift in the oscillation profile, with a maximum of approximately $\pi$, indicative of a Berry phase transition. Furthermore, the parameter regime in which anomalous oscillations are enhanced is further highlighted by a corresponding phase diagram, providing a basis for their experimental observation. Overall, our findings provide a clear understanding of how SOC influences anomalous quantum oscillations and offer guidance for the experimental observation of anomalous quantum oscillations.

\begin{acknowledgments}
This work was supported by the National Natural Science Foundation of China (Grant No.11874139), the Natural Science Foundation of Hebei Province (Grant No.A2019205190), and the Scientific Research Foundation of Hebei Normal University (Grant No.L2024J02).
\end{acknowledgments}

\appendix
%%%%%%%%%%%%%%%%%%%%%%%%%%%%%%%%%%%%%%%%%%%%%%%%
\section{Self-Energy Calculation Method}
\label{selfenergy}
Here, we present the numerical method for calculating the disorder-induced self-energy, as discussed in Eq.~\ref{eq3} of the main text. We employ the self-consistent Born approximation. The integral equation for the self-energy is given by:
\begin{equation}
\Sigma(E) = \frac{U^{2}}{12} \left(\frac{a}{2\pi}\right)^{2} \int_{\text{BZ}} d^{2}\mathbf{k} \left[\sigma_{0}(E^{+}I - H_{0} - \Sigma)^{-1}\sigma_{0}\right],
\label{eq:self_consistent_born}
\end{equation}
where \( \sigma_{0} \) is the identity matrix, \( U \) is the disorder strength, \( a \) is the lattice constant, and the integration is performed over the Brillouin zone (BZ). The superscript \( + \) denotes the retarded Green's function, with \( E^{+} = E + i0^{+} \).

The self-energy \( \Sigma \) is decomposed as:
\begin{equation}
\text{Re} \Sigma = \begin{pmatrix}
\bar{\mu} + \bar{M_0} & 0 \\
0 & \bar{\mu} - \bar{M_0}
\end{pmatrix},
\quad
\text{Im} \Sigma = \begin{pmatrix}
-\Gamma_{e} & 0 \\
0 & -\Gamma_{h}
\end{pmatrix},
\end{equation}
where \( \bar{\mu} \) and \( \bar{M_0} \) represent the renormalization of the chemical potential and the fundamental gap, respectively, and \( \Gamma_{e} \) and \( \Gamma_{h} \) are the scattering rates for the electron and hole bands, respectively. The self-energy matrix is in \(4 \times 4\) form, with the upper and lower \(2 \times 2\) blocks being identical.

%%%%%%%%%%%%%%%%%%%%%%%%%%%%%%%%%%%%%%%%%%%%%%%%
\section{Perturbation Treatment of the Spin-Orbit Coupling}
\label{SOC}

This section provides a detailed analysis of the energy corrections induced by spin-orbit coupling (SOC) due to structural inversion asymmetry (SIA). The SOC term is described by \( H_{\text{SIA}}(\mathbf{k}) \) as given in Eq.~(\ref{eq2}), and it is treated as a perturbation to the unperturbed Hamiltonian \( H_0(\mathbf{k}) \), which is defined in Eq.~(\ref{eq1}). For \( H_0(\mathbf{k}) \), the Hamiltonian is block-diagonal with doubly degenerate eigenvalues:
\begin{equation}
E^{\pm} = \frac{M_1 + M_2 \pm D}{2},
\end{equation}
where \( D = \sqrt{\delta^2 + 4\beta^2 k^2} \) and \( \delta = M_1 - M_2 \). The normalized components of the eigenstates are given by 
\( a_{\pm} = c_{\pm} = 1/\sqrt{1 + \eta_{\pm}} \), \( b_{\pm} = -(M_1 - E^{\pm})a_{\pm}/(\beta k_{+}) \), and \( d_{\pm} = (M_1 - E^{\pm})c_{\pm}/(\beta k_{-}) \), 
where \( \eta_{\pm} = (\delta \mp D)^2/(4\beta^2 k^2) \).
 
The perturbation \( V \equiv H_{\text{SIA}} \) has non-zero elements only at \( V_{13} = -i\alpha k_{-} \) and \( V_{31} = i\alpha k_{+} \). We first apply degenerate perturbation theory to calculate the first-order energy correction. For the \( E^{+} \) subspace, the perturbation matrix is:
\[
W^{+} = \begin{pmatrix} 0 & w \\ w^* & 0 \end{pmatrix}, \quad \text{where} \quad w = -i\alpha k_{-} a_{+} c_{+}.
\]
The eigenvalues of this Hermitian matrix are \( \lambda = \pm |w| \), where \( |w| = |\alpha| k \frac{4\beta^2 k^2}{4\beta^2 k^2 + (\delta - D)^2} \). This leads to the following energy splitting:
\begin{equation}
E_1^+ = E^{+} + |w|, \quad E_2^+ = E^{+} - |w|.
\end{equation}
Similarly, for the \( E^{-} \) subspace, the perturbation matrix is
\[
W^{-} = \begin{pmatrix} 0 & w_{-} \\ w_{-}^* & 0 \end{pmatrix}, \quad \text{where} \quad w_{-} = -i\alpha k_{-} a_{-} c_{-},
\]
with \(|w_{-}| = |\alpha| k \frac{4\beta^2 k^2}{4\beta^2 k^2 + (\delta + D)^2}.\) This also leads to the following energy splitting:
\begin{equation}
E_1^- = E^{-} + |w_{-}|, \quad E_2^- = E^{-} - |w_{-}|.
\end{equation}
Thus, after the first-order correction, the two degenerate levels split into four distinct levels: \(E_1^+, E_2^+, E_1^-, E_2^- \).

After the degeneracy is lifted by the first-order correction, we apply non-degenerate perturbation theory to calculate the second-order correction. The general formula for the second-order energy correction is:
\begin{equation}
E_n^{(2)} = \sum_{m \neq n} \frac{ |\langle m | V | n \rangle|^2 }{ E_n^{(0)} - E_m^{(0)} }.
\end{equation}
For a state \( |n\rangle \) from the \( E^{+} \) subspace, the significant contributions come from states \( |m\rangle \) in the \( E^{-} \) subspace. The sum over these intermediate states is independent of the specific state \( |n\rangle \):
\[
\sum_{m \in E^{-}} |\langle m | V | n \rangle|^2 = \alpha^2 k^2 \frac{1}{(1+\eta_{+})(1+\eta_{-})}.
\]
Using the identity \( (1+\eta_{+})(1+\eta_{-}) = D^2 / (\beta^2 k^2) \) and \( E^{+} - E^{-} = D \), the correction becomes:
\begin{equation}
E_n^{(2)} = \frac{ \alpha^2 k^2 }{ (1+\eta_{+})(1+\eta_{-}) D } = +\frac{ \alpha^2 \beta^2 k^4 }{ D^3 }.
\end{equation}
All states originating from the \( E^{+} \) subspace receive the same positive correction.

For a state \( |m\rangle \) originating from the \( E^{-} \) subspace, the intermediate states are from the \( E^{+} \) subspace, and a similar calculation yields:
\begin{equation}
E_m^{(2)} = -\frac{ \alpha^2 k^2 }{ (1+\eta_{+})(1+\eta_{-}) D } = -\frac{ \alpha^2 \beta^2 k^4 }{ D^3 }.
\end{equation}
All states originating from the \( E^{-} \) subspace receive the same negative correction.

In summary, the second-order correction depends only on the original unperturbed energy level. For states in the \( E^{+} \) subspace, the second-order correction is \( E^{(2)} = +\alpha^2 \beta^2 k^4 / D^3 \), while for states in the \( E^{-} \) subspace, the correction is \( E^{(2)} = -\alpha^2 \beta^2 k^4 / D^3 \). This correction is one order of magnitude smaller than the first-order energy shift and can therefore be safely neglected.

%%%%%%%%%%%%%%%%%%%%%%%%%%%%%%%%%%%%%%%%%%%%%%%%
\section{Landau Quantization under a Magnetic Field}
\label{magneticHamiltonian}

In this section, we present the explicit form of the magnetic Hamiltonian \( H_n \) for a given magnetic field \( B \). Starting from the Hamiltonian \( H = \mathbf{\Pi}^2/(2m) \) with the kinetic momentum operator \(\mathbf{\Pi} = \hbar\mathbf{k} + (e/c)\mathbf{A}\) in the Landau gauge \(\mathbf{A} = (-By, 0, 0)\), the key step is evaluating the commutator \([\Pi_x, \Pi_y] = -i\hbar e B / c\). This result motivates the introduction of the magnetic length \( l_B = \sqrt{\hbar c / (eB)} \) and the definition of dimensionless conjugate operators \(\hat{X} = (l_B/\hbar) \Pi_y\) and \(\hat{P} = (l_B/\hbar) \Pi_x\), which satisfy \([\hat{X}, \hat{P}] = i\). The Hamiltonian is thus isomorphic to a harmonic oscillator, allowing for the introduction of ladder operators \(\hat{a} = (\hat{X} + i\hat{P})/\sqrt{2}\) and \(\hat{a}^\dagger = (\hat{X} - i\hat{P})/\sqrt{2}\). The original wavevector operators are then expressed in terms of these ladder operators as \( k_{+} \rightarrow \sqrt{2} \hat{a}^{\dagger}/l_B \), \( k_{-} \rightarrow \sqrt{2} \hat{a}/l_B \), and \( k^{2} \rightarrow (2/l_B^2)(\hat{a}^{\dagger}\hat{a} + 1/2) \).

The eigenstates of the number operator \(\hat{N} = \hat{a}^\dagger \hat{a}\) form a complete basis for each Landau level, denoted by \(|n\rangle\), with the well-known properties \(\hat{a}^\dagger\hat{a}|n\rangle = n|n\rangle\), \(\hat{a}|n\rangle = \sqrt{n}|n-1\rangle\), and \(\hat{a}^\dagger|n\rangle = \sqrt{n+1}|n+1\rangle\). A generic Hamiltonian \(H(\mathbf{k})\), when expressed in terms of \(k_+\), \(k_-\), and \(k^2\), can be projected onto this basis. The action of the ladder operators ensures that the matrix representation of \(H\) becomes block-diagonal in the Landau level index \(n\), where each block \(H_n\) is a finite-dimensional matrix. The matrix elements are determined by applying the substitution rules for the wavevector operators and evaluating the action of \(\hat{a}\) and \(\hat{a}^\dagger\) on the states \(|n\rangle\), which yields the characteristic \(\sqrt{n}\) and \(\sqrt{n+1}\) factors.

For the specific model considered in the main text, this procedure leads to the effective Hamiltonian matrix \(H_n\) for a given Landau level index \(n\), as shown below.

\begin{widetext}
\begin{equation}
\begin{aligned}
 H_n = 
\begin{pmatrix}
M_0 + \dfrac{2\mu_{+}}{l_B^2}\left(n + \dfrac{1}{2}\right) -i \Gamma_e 
& \dfrac{\sqrt{2}\beta}{l_B}\sqrt{n} 
& -\mathrm{i}\dfrac{\sqrt{2}\alpha}{l_B}\sqrt{n + 1} 
& 0 
\\[8pt]
\dfrac{\sqrt{2}\beta}{l_B}\sqrt{n} 
& -M_0 - \dfrac{2\mu_{-}}{l_B^2}\left(n - \dfrac{1}{2}\right) -i \Gamma_h 
& 0 
& 0 
\\[8pt]
\mathrm{i}\dfrac{\sqrt{2}\alpha}{l_B}\sqrt{n + 1} 
& 0 
& M_0 + \dfrac{2\mu_{+}}{l_B^2}\left(n + \dfrac{3}{2}\right) -i \Gamma_e 
& -\dfrac{\sqrt{2}\beta}{l_B}\sqrt{n + 2} 
\\[8pt]
0 
& 0 
& -\dfrac{\sqrt{2}\beta}{l_B}\sqrt{n + 2} 
& -M_0 - \dfrac{2\mu_{-}}{l_B^2}\left(n + \dfrac{5}{2}\right) -i \Gamma_h
\end{pmatrix}
\end{aligned}
\label{eq:Hn_matrix}
\end{equation}
\end{widetext}

By solving the eigenvalue equation \( H_n|\Psi_n\rangle = E_n|\Psi_n\rangle \), we can obtain the energy levels \( E_n \) as a function of the Landau level index \( n \). These Landau levels are the fundamental input for calculating the density of states and other magnetic-field-dependent physical quantities discussed in the main text.

\section{Calculation of \texorpdfstring{$\delta A$}{delta A}}
\label{deltaA}
\begin{figure}
    \centering
    \includegraphics[width=0.45\textwidth]{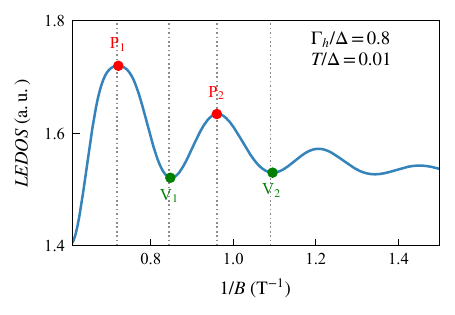}
    \caption{LEDOS as a function of inverse magnetic field $1/B$ for parameters $T = 0.01\Delta$ and $\Gamma_h = 0.8\Delta$. Characteristic peaks ($P_1$, $P_2$) and valleys ($V_1$, $V_2$) are identified, from which the oscillation amplitude $\delta A = (P_1 - V_1) + (P_2 - V_2)$ is calculated.}
    \label{subfig1}
\end{figure}
As an example, we consider the LEDOS oscillation curve at $T = 0.01\Delta$ and $\Gamma_h = 0.8\Delta$, shown in Fig.~\ref{subfig1}. In this oscillation curve, there are multiple peaks and valleys. The main oscillation peaks are labeled as $P_1$ and $P_2$, with the corresponding valleys $V_1$ and $V_2$. The oscillation amplitude $\delta A$ is calculated as
\begin{equation}
    \delta A = (P_1 - V_1) + (P_2 - V_2).
\end{equation}
This calculation provides a single data point in the phase diagram of Fig.~\ref{fig:7}(c), corresponding to $T = 0.01\Delta$ and $\Gamma_h = 0.8\Delta$.

\nocite{*}
\bibliographystyle{apsrev4-2}
\bibliography{ref}% Produces the bibliography via BibTeX.
% ****** End of file apssamp.tex ******
\end{document}